\documentstyle[12pt]{article}
\title{Non-Kolmogorov probability models and modified Bell's inequality}
\author{Andrei Khrennikov\\
Department of Mathematics, Statistics and Computer Sciences\\
University of V\"ax\"o, S-35195, Sweden}

\begin{document}
\maketitle

\begin{abstract} We analyse the proof of Bell's inequality and demonstrate that this inequality is 
related to one particular model of probability theory, namely Kolmogorov
measure-theoretical axiomatics, 1933. 
We found a (numerical) statistical correction to Bell's inequality. 
Such an additional term 
$\epsilon_\phi$ in the right hand side of Bell's inequality can be considered as 
a probability invariant of a quantum state $\phi.$ 
This is a measure of nonreproducibility of hidden variables in different runs 
of experiments. Experiments to verify Bell's inequality 
can be considered as just experiments to estimate the constant $\epsilon_\phi.$
It seems that  Bell's inequality could not be used as a crucial reason to deny
local realism. We consider deterministic as well as stochastic hidden variables
models.
\end{abstract}

{\bf 1. Introduction}
  

\medskip

Experimental violations [1] of Bell's inequality [2] are typically
(see, for example,  [1],[3]) interpreted 
in one of two ways: (1) {\bf nonlocality:}
by changing the  state of one particle in the EPR pair we change the state of the other 
particle; (2) {\bf death of reality:} realism could not be used as the philosophic 
base of quantum mechanics (`properties' of quantum systems are not objective properties,
i.e., the properties of an object). In particular, (2) implies that the statistical 
interpretation of quantum mechanics (via L. Ballentine, [4])
must be denied in favour of the orthodox Copenhagen interpretation. 
Although such a viewpoint is dominating in the quantum community,
there are still some doubts that violations of Bell's inequality
{\bf must be} interpreted in such a way. In particular, many scientists thought
(and continue to think) that  ``Bell's paradox" has purely probabilistic origin,
see, for example, [5]. Unfortunately these probabilistic considerations had merely
philosophic character. In any case they did not give a new (Bell-like) inequality
which has an experimental meaning. 

{\it Remark 1.1.}  The common opinion is that ``Bell's paradox" (experimental violations
of Bell's inequality) is just a reformulation of the EPR paradox. However, the problem
is more complicated: Bell's probabilistic reformulation of the EPR paradox contains
some additional assumptions (on probability distribution of hidden variables).

In this note we follow to the general 
(probabilistic) attitude of [5]. In fact, we generalize
ideas of De Baere, [5], on connection of Bell's inequality and ``implicit
reproducibility" (we arrived to such ideas independently 
by developing non-Kolmogorov probability formalisms, [6]).
However, we found some statistical
{\bf quantity} $\epsilon_\phi$ which can be considered as a probability invariant
of a quantum state $\phi.$ It seems that the standard interpretation of violations
of Bell's inequality is a consequence of neglecting of this quantity $\epsilon_\phi.$

We analyse Bell's proof and demonstrate that
the possibility to derive Bell's inequality depends crucially on the use of
a particular probabilistic model, namely the model based on Kolmogorov's axiomatics
[7], 1933 (so called measure-theoretical approach to probability).
Of course, the Kolmogorov model is the dominating mathematical model for probability
theory. Therefore it is not surprising that J. Bell and many others used 
this approach to probability theory. However, there also exist numerous non-Kolmogorov 
probabilistic models
(which are similar to non-Euclidean geometrical models), see, for example, [8] and
Accardi and  Gudder in [5].
In particular, I constructed probabilistic models, [6], which describe random 
phenomena in that the standard law of large numbers is violated: relative
frequencies $\nu_N= n/N$ have no limit (stabilization), $n\to \infty$ (numerous
examples of such a random behaviour can be found in [6]).

{\it Remark 1.2.} We want to underline that in physics the choice of the right probability model
is not less important then the choice of the right geometric model. The Kolmogorov
model (as well as Euclidean model) could not describe all physical phenomena.

In this note we analyse ``Bell's paradox"  on the basis of the assumption that
the law of large numbers can be violated for hidden variables: $\nu_N(\lambda)$
can fluctuate. From the physical viewpoint this means that different runs of experiments
(for example, for correlated particles) can produce different ``probability 
distributions" for hidden variables. In such a situation it would be impossible
to define a Kolmogorov probability distribution ${\bf P}$ on the set of hidden 
variables $\Lambda.$ Kolmogorov's model could not be applied. We introduce a numerical
measure for fluctuations $\epsilon_\phi$. 
It will be shown that ``general Bell's inequality" must contain
this probability invariant of a quantum state as an additional term.

Finally, we remark that all experimental calculations are, in fact, based not on Kolmogorov
model (probability as a measure), but on von Mises, 1919, [9] model (probability
as frequency). ``Experimental covariation" of two observables, $A,B,$
is calculated as a sequence mean value:
$<A, B>_{\rm{fr}}=(1/N) \sum_{i=1}^N A_i B_i,$ where $x=(A_1,A_2,..., A_N)$  and
$y=(B_1,B_2,..., B_N)$ are random sequences (collectives) generated by measurements 
of $A$ and $B.$  Hence, in fact, $<A, B>_{\rm{fr}}$ depends on $x$ and $y$: 
$<A, B>_{\rm{fr}}=<A,B>_{xy}.$ However, J. Bell supposed that there exists a 
Kolmogorov probability distribution ${\bf P}$ on the set of hidden variables 
$\Lambda$ and all covariations can be written as mean values with respect to this
unique measure: $< A, B>_{\rm{Bell/Kol}}= \int_\Lambda A(\lambda) B(\lambda) d {\bf P}(\lambda).$ 
This (rather strong)  assumption statistical postulate has  never been verified experimentally.

\medskip

{\bf 2. Bell's proof}

\medskip

We reproduce the proof of Bell's inequality. 
Let  ${\cal P}=(\Omega, {\cal F}, {\bf P})$
be a Kolmogorov probability space: $\Omega$ is a space of elementary events,
${\cal F}$ is an algebra of events, ${\bf P}$ probability measure.

{\bf Theorem 1.} {\it Let  
$A,B,C= \pm 1$ be random variables on ${\cal P}.$  Then Bell's inequality 
\begin{equation}
\label{BBB}
\vert <A,B > - < C,B >\vert \leq 1 - <A,C>
\end{equation}
holds true.}

{\bf Proof.} Set $\Delta= <A,B > - < C,B >.$ By linearity of Lebesgue
integral we obtain
\begin{equation}
\label{B1}
\Delta = \int_\Omega A(\omega) B(\omega) d {\bf P}(\omega)- 
\int_\Omega C(\omega) B(\omega) d {\bf P}(\omega)=
\int_\Omega [A(\omega) - C(\omega)]B(\omega) d {\bf P}(\omega).
\end{equation}
As $A(\omega)^2= 1,$ 
\begin{equation}
\label{B2}
\vert \Delta\vert = 
\vert \int_\Omega [1 - A(\omega) C(\omega)] A(\omega) B(\omega) d {\bf P}(\omega)\vert
\leq \int_\Omega [1 - A(\omega) C(\omega)]  d {\bf P}(\omega).
\end{equation}
 
{\small Of course, this is the rigorous mathematical proof of (\ref{BBB}) for 
Kolmogorov probabilities. However, abstractness of Kolmogorov's probability model
induces serious problems, if we do not control carefully dependence of
probabilities on  corresponding statistical  ensembles of physical systems. Bell
did not control this dependence. In fact, the symbol ${\bf P}$ of probability
which is used in the proof 
must be regarded to different statistical ensembles.}

\medskip

{\bf 3. Fluctuating distributions of hidden variables}

\medskip

To simplify our considerations, we suppose that the set of hidden variables is finite:
$
\Lambda=\{ \lambda_1,...,  \lambda_M\}.
$
For each physical observable $U$, the value $\lambda$ of hidden variables determines the
value 
$
U=U( \lambda).
$
Let $U$ and $V$ be physical observables, $U,V=\pm 1.$ We start with the 
consideration of the frequency
(experimental) covariation $< U, V>_{x_{UV}}$ with respect to a random sequence  
$x_{UV}= (x_1,x_2,..., x_N,...),$ where $x_i= (u_i, v_i),$ which is induced 
by measurements of the pair $(U,V).$ The $x_{UV}$ is obtained by measurements for
an ensemble $S_{UV}$ of physical systems 
(for example, pairs of correlated 
quantum particles). Our aim is to represent experimental covariation
$< U, V>_{x_{UV}}$ as ensemble covariation $< U, V>_{S_{UV}}.$ Then we shall demonstrate
that in the general case it is impossible to perform for ensemble covariations
Bell's calculations which have been performed for Kolmogorov 
covariations. Let $S_{UV}=\{d_1,..., d_N\},$ where  
$i$th measurement is performed for the system $d_i.$ 
Define a function $i \to \lambda(i),$  the value of hidden variables for $d_i.$
We set $n_k(S_{UV})= \vert \{ d_i\in S_{UV} : \lambda(i) = \lambda_k\} \vert$ and
${\bf p}_k^{UV} = {\bf P}_{S_{UV}}(\lambda= \lambda_k)=\frac{n_k(S_{UV})}{N}.$
These are probabilities of hidden variables $\lambda_k, k=1,2,...,M,$ in the statistical
ensemble $S_{UV}.$ We have
$
< U, V>_{x_{UV}}= \frac{1}{N} \sum_{i=1}^N U(\lambda(i)) V(\lambda(i))
= \sum_{k=1}^M {\bf p}_k^{UV} u_k v_k =<U,V>_{S_{UV}},
$
where $u_k =U(\lambda_k), v_k =V(\lambda_k).$ 
Thus 
$$
\Delta = <A,B>_{x_{AB}} - <C,B>_{x_{CB}} 
$$
$$
=
<A,B>_{S_{AB}} - <C,B>_{S_{CB}} =
\sum_k ( {\bf p}_k^{AB} a_k - {\bf p}_k^{CB} c_k) b_k
$$
and
$$
<A,C>_{x_{AC}} = <A,C>_{S_{AC}}=
\sum_k {\bf p}_k^{AC} a_k c_k.
$$
We now suppose  that 
{\it probabilities of $\lambda_k$ do not depend on statistical ensembles:}
\begin{equation}
\label{L}
{\bf p}_k ={\bf p}_k^{AB}= {\bf p}_k^{CB}= {\bf p}_k^{AC}
\end{equation}
(later we shall modify this condition to obtain statistical 
coincidence of probabilities, instead of the precise coincidence).
Hence
$
\Delta=
\sum_{k=1}^M {\bf p}_k (a_k - c_k) b_k\; \; \mbox{and}\; \;
<A,C>_{x_{AC}}=\sum_{k=1}^M {\bf p}_k a_k c_k .
$
We can now apply Theorem 1 for the discrete probability distribution 
$\{ {\bf p}_k \}_{k=1}^M$ and obtain Bell's inequality.

However, if condition (\ref{L}) does not hold true, then equality
(\ref{B1}) and, as a consequence, Bell's inequality can be violated.
The violation of condition (\ref{L}) is the exhibition of unstable
statistical structure on the level of hidden variables. Condition
(\ref{L}) is equivalent to a condition of implicit reproducibility
which was discussed by De Baere [5].

{\it Remark 3.1.} ($p$-adic probability models, negative probabilities 
and Bell's inequality). All our considerations
were based on the statistical stabilization with respect to the real metric. In [6]
we  considered the statistical stabilization with respect to a $p$-adic metric.
The field
of $p$-adic numbers ${\bf Q}_p,$ where $p>1$ is a prime number, can be constructed
(as the field of real numbers ${\bf R}$) as a completion of the field of rational
numbers ${\bf Q}.$ The $p$-adic metric differs strongly from the real one.
As for finite ensembles $S,$ ensemble probabilities ${\bf P}_S(a) = \frac{n(a)}{N}$
are rational numbers, we can study their behaviour not only with respect to the real metric
on ${\bf Q},$ but also with respect to the $p$-adic metric.
$p$-adic probability theory gives numerous examples of ensemble probabilities
fluctuating in the real metric and stabilizing in the $p$-adic metric. However,
the $p$-adic stabilization of probabilities does not imply the possibility to repeat
Bell's proof for $p$-adic probabilities: these probabilities may be negative rational 
numbers, see [6] (compare with Muckenheim, [5]).

\medskip

{\bf 4. Measure of statistical deviation between runs of an experiment}

\medskip

We introduce now a statistical analogue of the precise coincidence of ensemble probabilities
for hidden variables.
Let ${\cal E}_1, {\cal E}_2$ be two ensembles of physical systems
and let $\pi$ be a property of elements of these ensembles. The $\pi$
has values $(\alpha_1,..., \alpha_m).$ We define
$$
\delta_\pi ({\cal E}_1, {\cal E}_2) = \sum_{i=1}^M \vert {\bf P}_{{\cal E}_1} (\alpha_i) -
{\bf P}_{{\cal E}_2} (\alpha_i)\vert,
$$
where ${\bf P}_{{\cal E}} (\alpha_i)= \frac{\vert \{d \in {\cal E}: \pi(d)= \alpha_i\}\vert}
{\vert {\cal E}\vert}$ are ensemble probabilities. We remark that the function $\delta=\delta_\pi$
is a pseudometric on the set of all ensembles which elements have the property $\pi:
\; 1) \delta({\cal E}_1, {\cal E}_2) \geq 0; 2) \delta ({\cal E}_1, {\cal E}_2)
=\delta ({\cal E}_2, {\cal E}_1); 3) \delta ({\cal E}_1, {\cal E}_2)\leq
\delta ({\cal E}_1, {\cal E}_3) + \delta ({\cal E}_3, {\cal E}_2).$
In our model we set $\pi= \lambda,$ hidden variables. The precise reproducibility of the 
probability distribution of hidden variables (\ref{L}) can be written as
$$
\delta(S_{AB}, S_{CB})= \delta(S_{AB}, S_{AC})=0,
$$
where $\delta= \delta_\lambda.$ Of course, we need not use such a precise coincidence
in probabilistic considerations. Let $\phi$ be a quantum state. Denote 
by the symbol $T_\phi$
the set of all statistical ensembles ${\cal E}$ which correspond to $\phi$
(can be obtained with the aid of some preparation 
procedure corresponding to $\phi).$ Set 
$$
\epsilon_\phi=
\sup\{ \delta({\cal E}_1, {\cal E}_2): {\cal E}_1, {\cal E}_2 \in T_\phi\}.
$$

{\bf Theorem 2.} (``General Bell's inequality")
{\it Let $\phi$ be a quantum state and let $A, B, C$ be physical observables
such that pairs of observables $(A,B), (C,B)$ and $(A,C)$ can be measured.
Then inequality
\begin{equation}
\label{BBB1}
\vert <A,B > - < C,B > \vert \leq (1+2 \epsilon_\phi) - <A,C>
\end{equation}
holds true.}

{\bf Proof.} We have 
$$
\vert \Delta\vert = \vert <A,B>_{x_{AB}} - <C,B>_{x_{CB}} \vert 
$$
$$
\leq 
\vert \sum_{k=1}^M {\bf p}_k^{AB} (a_k - c_k) b_k\vert +
\vert \sum_{k=1}^M ({\bf p}_k^{AB} - {\bf p}_k^{CB}) c_k b_k\vert
$$
$$
\leq \epsilon_\phi + 
\sum_{k=1}^M {\bf p}_k^{AB} \vert a_k b_k\vert (1 -  a_k c_k) \leq
(1+\epsilon_\phi) - <A,C>_{S_{AC}} + 
 \sum_{k=1}^M \vert {\bf p}_k^{AC} - {\bf p}_k^{AB}\vert \vert a_k c_k \vert
$$
$$
\leq (1+ 2 \epsilon_\phi) - <A,C>_{S_{AC}}.
$$
\hfill\rule{2mm}{2mm} 

We use the index $N$ to denote the cardinality of a statistical ensemble.
If probabilities ${\bf P}_{S_{UV}^N}(\lambda_k)$ stabilize when $N\to \infty,$
$$
\lim_{N \to \infty} {\bf P}_{ S_{UV}^N}(\lambda_k)= {\bf P}(\lambda_k),
$$
then $\epsilon_\phi^N \to 0, N \to \infty.$
This imply precise Bell's inequality (\ref{BBB}).

Experiments to verify  Bell's inequality can be considered as experiments 
to estimate the probability invariant $\epsilon_\phi$ for some class of 
quantum states. It seems that the only lesson of these experiments is that
{\it there exist quantum states $\phi$ which have nonzero probability invariant
$\epsilon_\phi.$} It may be that physical reality is nonlocal. It may be that
it is even nonreal. However, it seems that Bell's arguments did not 
imply neither nonlocality nor nonreality.

\medskip

{\bf  5. Stochastic hidden variables model} 

\medskip

In this section it is supposed that the result of a measurement depends not only on the value
$\lambda$ of hidden variables, but also on the state $\omega^U$ of an 
equipment ${\cal M}_U$ which is used for
measuring of $U.$ This the empiricists (contextualistic realists)
interpretation of quantum mechanics, see, for example,  
 W. De Muynck, W. De Baere, H. Marten, Ref. [5].

A measurement device ${\cal M}_U$ is a complex macroscopic system
which state depends on the huge number of fluctuating parameters. Denote the ensemble
of all possible states of ${\cal M}_U$ by the symbol $\Sigma_U: \;
\Sigma_U= \{ \omega_1^U,...,  \omega_{L_U}^U \}.$ The final value $U_f$ of an observable
$u$ depends on both  $\lambda$ and $\omega:$
$$
u=U(\omega, \lambda).
$$
We call such a model {\it stochastic hidden variables model.} Our model of
stochastic hidden variables  differs from the standard one, see section 6. 
The latter model  is strongly connected with Kolmogorov's probability model
(existence of the probability distribution of hidden variables ${\bf P}(\lambda)$
and conditional probabilities ${\bf P}(U, \lambda)$ is postulated).

Let $U$ and $V$ be physical observables, $U, V=\pm 1.$ We start again with 
the consideration of the frequency
covariation $< U, V>_{x_{UV}}$ with respect to a collective  $x_{UV}$ induced 
by the measurement of the pair $(U,V).$ The $x_{UV}$ is obtained by measurements for
an ensemble $S_{UV}$ of physical systems. 
Our aim is again to represent the experimental covariation
$< U, V>_{x_{UV}}$ as ensemble covariation $< U, V>_{S_{UV}}.$ Then we shall demonstrate
that in the general case it is impossible to perform for ensemble covariations
Bell's calculations, (\ref{B1}) -- (\ref{B2}).

Let $S_{UV}=\{d_1,..., d_N\},$ where  $i$th measurement is performed for the system $d_i.$ 
Define functions $i\to \lambda(i)$ (the same function as above) and $i \to \omega^U(i),
i \to \omega^V(i),$ states of apparatus ${\cal M}_U$ and ${\cal M}_V,$ respectively,
at the instances, $t_i^U$ and $t_i^V,$ 
of measurements of $U$ and $V$ for $i$th system.  We have
$$
< U, V>_{x_{UV}}= \frac{1}{N} \sum_{i=1}^N U(\omega^U(i), \lambda(i))
V(\omega^V(i), \lambda(i)).
$$
Set $D_{ks}^U =\{ i : \lambda(i) = \lambda_k,
\omega^U(i) =\omega^U_s\}$ and $D_{ks}^V =\{ i : \lambda(i) = \lambda_k,
\omega^V(i) =\omega^V_s\}, 1\leq k \leq M, 1\leq s\leq L_U,
1\leq q\leq L_V.$ 
Set $l_{ksq}^{UV} = \vert D_{ks}^U\cap D_{kq}^V\vert.$ It is evident that 
$$
\sum_{k=1}^M \sum_{s=1}^{L_U } \sum_{q=1}^{L_V }l_{ksq}^{UV} =N.
$$
Hence
$$
< U, V>_{x_{UV}}= \frac{1}{N} \sum_{k s q} l_{ksq}^{UV} u_{ks} v_{kq},
$$
where $u_{ks}= U(\omega_s^U,\lambda_k),  v_{kq}= V(\omega_q^V,\lambda_k).$
We show that $< U, V>_{x_{UV}}$ can be represented as ensemble covariation for
an appropriative ensemble of physical systems and states of measurement devices. 
However, a choice of such an ensemble is rather delicate problem.

First we
note that $< U, V>_{x_{UV}}\not = < U, V>_{\Lambda\times \Sigma_A\times \Sigma_B}$
(compare with section 6).
For the latter covariation, we have
$$
< U, V>_{\Lambda\times \Sigma_A\times \Sigma_B}=\frac{1}{M L_A L_B}
\sum_{k=1}^M \sum_{s=1}^{L_U } \sum_{q=1}^{L_V }u_{ks} v_{kq}
$$
and in general ${\bf P}_{\Lambda\times \Sigma_A\times \Sigma_B} (\lambda=\lambda_k,
\omega^U = \omega^U_s, \omega^V = \omega^V_q) = \frac{1}{M L_A L_B} \not=
\frac{l_{ksq}}{N}$ even approximately for $M, N, L_A,L_B  \to \infty.$

It is also evident that $< U, V>_{x_{UV}}\not = < U, V>_{S_{UV}}.$ The latter covariation
is simply not well defined, because the `properties'
$\omega^U(i) = \omega^U_s, \omega^V(i) = \omega^V_q$ are not objective properties of elements
of the ensemble $S_{UV}.$ These `properties' are determined by fluctuations of parameters 
in the apparatus ${\cal M}_U$ and ${\cal M}_V.$ 

To find the right ensemble, we have to introduce two new ensembles, namely, ensembles of states
of the apparatus ${\cal M}_U$ and ${\cal M}_V$ (in the process of measurements for 
the ensemble of physical systems $S_{UV}):$ 
$$
S_{{\cal M}_U}=\{ \alpha_1^U,..., \alpha_N^U\}, \;\alpha_j^U \in \Sigma_U, \;\;
S_{{\cal M}_V}=\{ \alpha_1^V,..., \alpha_N^V\}, ,\alpha_j^V \in \Sigma_V ,
$$
where $\alpha_i^U= \omega^U(i), \alpha_i^V=\omega^V(i)$ 
are states of ${\cal M}_U$ and ${\cal M}_V$
at the instances of $i$th measurements. We set
$$
{\bf S}_{UV} = \rm{diag}(S_{UV} \times S_{{\cal M}_U} \times S_{{\cal M}_V})
= \{ D_1,..., D_N \}, \; \; D_j = (d_j, \alpha_j^U,  \alpha_j^V).
$$
Then $\pi(D_j)= (\lambda(j), \omega^U(j),  \omega^V(j))$ is an objective property of elements
of the ensemble ${\bf S}_{UV}$ and
$$
< U, V>_{x_{UV}}= < U, V>_{{\bf S}_{UV}}= \frac{1}{N} \sum_{i=1}^N U(\omega^U(i), \lambda(i))
V(\omega^V(i), \lambda(i)).
$$
We set 
$$
{\bf p}_{ksq}^{UV}= {\bf P}_{{\bf S}_{UV}}(D_j: \pi(D_j)= (\lambda_k, \omega_s^U,
\omega_s^V))
$$
$$
= \frac{\vert \{ D_j\in {\bf S}_{UV}: \pi(D_j)= (\lambda_k, \omega_s^U,
\omega_s^V)\}\vert}{\vert {\bf S}_{UV}\vert}.
$$
Hence we obtained that
$$
< U, V>_{x_{UV}}= < U, V>_{{\bf S}_{UV}}=
\sum_{k s q} {\bf p}_{ksq}^{UV}\; u_{ks} v_{kq}.
$$
Thus in the general case we have
$$
\begin{array}{rl}
&
\Delta= < A, B>_{x_{AB}} - < C, B>_{x_{CB}}=
< A, B>_{{\bf S}_{AB}} - < C, B>_{{\bf S}_{CB}} 
\\ & \\
&
=
\sum_{k s q} {\bf p}_{ksq}^{AB} \; a_{sk} b_{kq}
- \sum_{k s q} {\bf p}_{ksq}^{CB} \; c_{ks} b_{kq}
\\
\end{array}
$$
and
$$
< A, C>_{x_{AC}}=< A, C>_{{\bf S}_{AC}}=
\sum_{k s q} {\bf p}_{ksq}^{AC} \;a_{ks} c_{kq}.
$$
We suppose now that probabilities 
${\bf p}_{ksq}^{UV}$ do not depend on ensembles:
\begin{equation}
\label{S1}
{\bf p}_{ksq}= {\bf p}_{ksq}^{AB}= {\bf p}_{ksq}^{CB}= {\bf p}_{ksq}^{AC}. 
\end{equation}
In particular, we suppose that all measurement devices have the same set of states
(of parameters):
\begin{equation}
\label{S2}
\Sigma= \Sigma_A=\Sigma_B=\Sigma_C \; \; (\mbox{and}\; \; L= L_A =L_B= L_C). 
\end{equation}

Then we obtain
$$
\Delta=
\sum_{k s q} {\bf p}_{ksq} (a_{ks}- c_{ks}) b_{kq}.
$$
However, we could not 
repeat trick (\ref{B2}) of the proof of Bell's inequality. The equality 
$a_{ks}^2 = 1$ does not give the possibility to proceed the proof. Of course, we have
$$
\begin{array}{rl}
&
\vert \Delta \vert=
\vert \sum_{k s q} {\bf p}_{ksq} (a_{ks}- a_{ks}^2 c_{ks}) b_{kq}\vert
\leq \sum_{k s q} {\bf p}_{ksq} \vert a_{ks} b_{kq}\vert
(1- a_{ks} c_{ks}) 
\\ & \\
&
\leq 1 - \sum_{k s q} {\bf p}_{ksq} a_{ks} c_{ks}.
\\
\end{array}
$$
But in general $\sum_{k s q} {\bf p}_{ksq} a_{ks} c_{ks}$ is not larger
than $<A,C>_{x_{AC}}=\sum_{k s q} {\bf p}_{ksq} a_{ks} c_{kq}.$ 

Therefore, if we keep to empiricism, then even stability condition 
(\ref{S1}) (for combined ensembles of physical systems and states of measurement
apparatus)
 does not imply Bell's inequality. A new source of violation of Bell's inequality is 
the {\it inconsistency} of random fluctuations for two measurement devices
${\cal M}_U$ and ${\cal M}_V.$ In general $\omega^U(i) \not= \omega^V(i).$ 

Suppose that it could  be possible to
control states of ${\cal M}_U$ and ${\cal M}_V$ and choose $\omega$ for 
${\cal M}_U$ and ${\cal M}_V$ in the consistence way:
$$
\omega= \omega^U(i) =\omega^V(i).
$$
Then the ensemble ${\bf S}_{UV}$ would contain only triples of the form
$(\lambda_k, \omega_s,  \omega_s)$ and 
\begin{equation}
\label{S3}
{\bf p}_{ksq}^{UV}= {\bf P}_{{\bf S}_{UV}}(\lambda_k, \omega_s^U, 
\omega_q^V) = 0, \; s\not=q. 
\end{equation}
In such a case we obtain covariations:
$$
<U,V>_{\rm{Ideal}} = 
\frac{1}{N} \sum_{i=1}^N U(\omega^U(i), \lambda(i))
V(\omega^V(i), \lambda(i))= 
 \sum_{ks} {\bf p}_{ks}^{UV} u_{ks} v_{ks},
$$
where ${\bf p}_{ks}^{UV}= {\bf p}_{kss}^{UV}.$
If we also suppose the validity of (\ref{S1}), we obtain
$$
\vert \Delta_{\rm{Ideal}}\vert =
\vert \sum_{ks} {\bf p}_{ks} (a_{ks} - c_{ks})b_{ks} \vert
$$
$$
\leq 1 - \sum_{ks} {\bf p}_{ks} a_{ks}c_{ks}
=1 - <A,C>_{\rm{Ideal}}.
$$
However, ideal covariations have no direct connection to experimental frequency 
covariations.

Nevertheless, we can formulate the following mathematical theorem:

{\bf Theorem 3.} {\it Let statistical ensembles (physical systems/measurement apparatus)
satisfy conditions (\ref{S1}) and (\ref{S3}). Then Bell's inequality (\ref{BBB})
holds true for covariations with respect to these ensembles.}

Therefore, to obtain Bell's inequality in the empiricists framework, we have to suppose:
(1) statistical repeatability of ensemble distribution of hidden variables 
$\lambda$ in ensembles which are used for measurements;
(3) statistical repeatability of fluctuations of states $\omega$ in ensembles of an
equipment; (3) consistency of fluctuations of all measurement devices. 

If the reader even deny the possibility of violations of (1) or (2), he must agree that condition (3)
seems to be nonphysical: we could never control fluctuations of the huge number of parameters in the 
equipment.

Instead of precise coincidence (\ref{S1}), it is possible to consider
(under the assumption (\ref{S2}))
the statistical coincidence based on the quantity:
$$
\delta({\bf S}_{AB}, {\bf S}_{CB}) =
\sum_{k=1}^M \sum_{s=1}^L \sum_{q=1}^L 
\vert {\bf p}_{ksq}^{AB} - {\bf p}_{ksq}^{CB}\vert.
$$
Here  $\delta=\delta_\pi$ for
the property $\pi(i)= (\lambda(i), \omega^U(i), \omega^V(i)).$
We remark that condition (\ref{S1}) of the precise coincidence can be written as
$$
\delta({\bf S}_{AB}, {\bf S}_{CB})=0 
$$
for every two pairs of observable $(A,B)$ and $(C,B).$
We  also introduce a new quantity which is a statistical measure
of inconsistency of ensembles $S_{{\cal M}_U}$ and $S_{{\cal M}_V}:$
$$
\sigma({\bf S}_{UV})= \sum_{s\not=q} {\bf P}_{{\bf S}_{UV}} (\omega^U = \omega_s,
\omega^V = \omega_q)= \sum_k \sum_{s\not=q} {\bf p}_{ksq}^{UV}.
$$
Condition (\ref{S3}) of the precise consistency for states of ${\cal M}_U$
and ${\cal M}_V$ can be written in the form:
$$
\sigma({\bf S}_{UV})=0.
$$

{\bf Theorem 4.} {\it Let statistical ensembles
(physical systems/measurement apparatus) satisfy conditions:
$$
\delta({\bf S}_{AB}, {\bf S}_{CB}), \delta({\bf S}_{AB}, {\bf S}_{AC}) \leq \epsilon
\; \; \mbox{and}\; \;
\sigma({\bf S}_{AB}), \sigma({\bf S}_{CB}), \sigma({\bf S}_{AC})\leq \epsilon^\prime.
$$
Then inequality
$$
\vert <A,B>_{{\bf S}_{AB}}  -  <C,B>_{{\bf S}_{CB}}\vert \leq
(1+ 2 \epsilon+ 3  \epsilon^\prime) - <A,C>_{{\bf S}_{AC}}
$$
holds true.}

{\bf Proof.} We have 
$$
\vert \Delta\vert \leq
\epsilon+ \vert
 \sum_{k s q} {\bf p}_{ksq}^{AB} (a_{ks}- c_{ks}) b_{kq}\vert
$$
$$
\leq \epsilon+ 2 \epsilon^\prime+
\sum_{k s} {\bf p}_{ks}^{AB}  \vert (a_{ks}- c_{ks}) b_{ks}\vert
\leq 
\epsilon+ 2 \epsilon^\prime
+ \sum_{k s} {\bf p}_{ks}^{AB} (1- a_{ks}c_{ks})
$$
$$
\leq \epsilon+ 4 \epsilon^\prime
+ \sum_{k s q} {\bf p}_{ksq}^{AB} (1- a_{ks}c_{kq})
\leq (1+2\epsilon+ 4 \epsilon^\prime) - 
\sum_{k s q} {\bf p}_{ksq}^{AC} a_{ks}c_{kq}.
$$
\hfill\rule{2mm}{2mm} 

We remark again that experiments to test Bell's inequality
can be interpreted as just experiments to find an estimate
for a constant $C= 2\epsilon+ 4 \epsilon^\prime.$ From this point of view
the only result of these experiments is that $C$ is essentially larger that 
zero. However, such a results could be expected: it would be rather
strange if measures of statistical deviations $\delta$ and $\sigma$ would be equal
to zero despite of  fluctuations of parameters of measuring devices.

\medskip

{\bf 6. Probability distributions in
stochastic hidden variables models.} 

\medskip

Typically 
stochastic hidden variables models are defined as models with probabilities
$(\epsilon=\pm 1)$
\begin{equation}
\label{S4}
{\bf P}(U=\epsilon)=\int_{\Lambda} {\bf P}(U=\epsilon/ \lambda) d \; \rho( \lambda),
\end{equation}
where $\rho(\lambda)$ is the probability distribution of hidden variables and 
${\bf P}(U=\epsilon/ \lambda)$ is the conditional probability to measure the value
$U=\epsilon$ for the quantum system having the hidden state $\lambda,$
see, for example, [10]. 

Then (see Ref. [11]) the joint 
probability distribution can be defined (at least mathematically) as
\begin{equation}
\label{S5}
{\bf P}(U_1=\epsilon_1, U_2=\epsilon_2, U_3=\epsilon_3)=
\int_{\Lambda} {\bf P}(U_1=\epsilon_1/ \lambda)
{\bf P}(U_2=\epsilon_2/ \lambda){\bf P}(U_3 =\epsilon_3/ \lambda) d \; \rho( \lambda).
\end{equation}
In fact, to derive Bell's inequality in the Kolmogorov framework, it is sufficient
to use the existence (on the mathematical level) of the joint probability
distribution (\ref{S5}). However, considerations in the framework of the ensemble probability theory
demonstrated that `probabilities' (\ref{S4}) 
has no physical meaning. These are probabilities
with respect to the ensemble $\Lambda\times \Sigma_U.$ However, physical probabilities
are probabilities with respect to the ensemble ${\bf S}_U= \rm{diag} 
(S_U\times S_{{\cal M}_U}),$ where $S_U=\{d_1,...,d_N\}$ is the ensemble of quantum system
used in the measurement. We note that physical arguments against existing
of representation (\ref{S4}) were presented  by  W. De Muynck, W. De Baere, H. Marten
in Ref. [5].  I think that results of this paper can strongly improve
their considerations. We hope that our numerical description of nonexistence
of Kolmogorov probabilities could essentially clarify the problem.

\medskip

{\bf 6. Other probabilistic models which do not contradict to local realism.}

\medskip

L. Accardi in Ref. [5] used non-Kolmogorov model without Bayes' formula
to eliminate Bell's inequality from considerations related to spin's model.
Recently he developed a new  model which gives
an explanation of violations of Bell's inequality, see Ref. [12]. In fact, to get
``physical Bell's inequality" we have to consider in Theorem 1 indexed 
random variables $U^1$ and $U^2$ corresponding to correlated particles, 1 and 2.
``Physical Bell's inequality" can be obtained only on the basis of the implicit
anticorrelation: $U^1= - U^2.$ Accardi discussed the role of this condition
in Bell's arguments.

I. Pitowsky in Ref. [5]  discussed the possibility that some 
nonmeasurable sets can be physical events, i.e, some physical observables may be 
nonmeasurable.
There is no Bell's inequality in this approach. Thus there is no problem
with violations of Bell's inequality.
This model is consistent with known 
polarization phenomena and the existence of macroscopic magnetism. He also proposed a
thought experiment which indicates a deviation from the predictions of quantum mechanics.
We note that already A. N. Kolmogorov discussed 
`generalized probabilities' on the algebra  of all subsets of $\Omega.$
Mathematicians, in particular applied mathematicians, where reluctant to
take nonmeasurable sets seriously. As a result there was no mathematical
theory that relates nonmeasurable distributions with relative frequencies.
Such an extension of probability theory was created by I. Pitowsky and then strongly 
mathematically improved by S.P. Gudder in Ref. [5]. He introduced 
the concept of a probability manifold $M.$ The global
properties of $M$ inherited from its local structure were then considered.
It was shown that a deterministic spin model due to Pitowski falls within
this general framework. Finally, Gudder constructed a phase-space model for
nonrelativistic quantum mechanics. These two models give the same global
description as conventional quantum mechanics. However, they also give
a local descriptions which is not possible in conventional quantum
mechanics.

\bigskip

{\bf ACKNOWLEDGMENTS}

This paper was completed during the
visit to Moscow State University on the basis of the
grant of the  Royal Academy of Science (Sweden) for the collaboration
with the former Soviet Union.

{\bf References}

[1] J.F. Clauser ,  A. Shimony, {\it Rep. Progr.Phys.,} 
{\bf 41} 1881-1901 (1978).
 A. Aspect,  J. Dalibard,  G. Roger, 
{\it Phys. Rev. Lett.}, {\bf 49}, 1804-1807 (1982);
 D. Home,  F. Selleri, {\it Nuovo Cim. Rivista,} {\bf 14},
2--176 (1991).

[2] J.S. Bell,
{\it  Rev. Mod. Phys.}, {\bf 38}, 447--452 (1966).

[3] B. d'Espagnat, {\it Veiled Reality. An anlysis of present-day
quantum mechanical concepts.} Addison-Wesley(1995).

[4]  L. E. Ballentine, {\it Rev. Mod. Phys.}, {\bf 42}, 358--381 (1970). 

[5]  L. De Broglie, {\it La Physique Quantique Restera--t--elle 
Indeterministe?} Gauthier-Villars, Paris (1953);
Lochak G., De Broglie's initial conception of de Broglie waves.
{\it The wave--particle dualism. A tribute to Louis de Broglie on his 90th 
Birthday}, Edited by S. Diner, D. Fargue, G. Lochak and F. Selleri.
D. Reidel Publ. Company, Dordrecht, 1--25 (1970);
Accardi  L., The probabilistic roots of the quantum mechanical 
paradoxes. {\it Ibid}, 47--55;
W. De Muynck and W. De Baere W.,
{\it Ann. Israel Phys. Soc.}, {\bf 12}, 1-22 (1996);
 W. De Muynck, W. De Baere, H. Marten,
{\it Found. of Physics}, {\bf 24}, 1589--1663 (1994);
W. De Baere, {\it Lett. Nuovo Cimento}, {\bf 39}, 234-238 (1984);
I. Pitowsky, {\it Phys. Rev. Lett}, {\bf 48}, N.10, 1299-1302 (1982);
{\it Phys. Rev. D}, {\bf 27}, N.10, 2316-2326 (1983);
 S.P. Gudder, {\it J. Math Phys.,}
{\bf 25}, 2397- 2401 (1984);
W. Muckenheim, {\it Phys. Reports,} {\bf 133}, 338--401 (1986).

[6]. A. Yu. Khrennikov, {\it Dokl. Akad. Nauk SSSR, ser. Matem.},
 {\bf 322}, No. 6, 1075--1079 (1992);{\it J. Math. Phys.}, {\bf 32}, No. 4, 932--937 (1991);
{\it Physics Letters A}, {\bf 200}, 119--223 (1995);
{\it Physica A}, {\bf 215}, 577--587 (1995); {\it  Int. J. Theor. Phys.}, {\bf 34},
2423--2434 (1995); {\it J. Math. Phys.}, {\bf 36},
No.12, 6625--6632 (1995);
A.Yu. Khrennikov, {\it $p$-adic valued distributions in 
mathematical physics.} Kluwer Academic Publishers, Dordrecht (1994);
A.Yu. Khrennikov, {\it Non-Archimedean analysis: quantum
paradoxes, dynamical systems and biological models.}
Kluwer Acad.Publ., Dordreht, The Netherlands, 1997.

[7] Kolmogoroff A. N., {\it Grundbegriffe der Wahrscheinlichkeitsrechnung.}
Springer Verlag, Berlin (1933); reprinted:
{\it Foundations of the Probability Theory}. 
Chelsea Publ. Comp., New York (1956).

[8]  T. L. Fine, {\it Theories of probabilities, an examination of
foundations}. Academic Press, New York (1973). 

[9]  R.  von Mises, {\it The mathematical theory of probability and
 statistics}. Academic, London (1964).

[10] P.H. Eberhard, {\it Nuovo Cimento}, B, {\bf 46}, 392-400 (1978);
W. de Muynck, J. Steklenborg, {\it Ann. Phys.}, {\bf 45}, 222-234(1988).

[11] A. Fine, {\it Phys. Rev. Letters}, {\bf 48}, 291--295 (1982).

\end{document}